\documentclass[%
 reprint,
 superscriptaddress,
 frontmatterverbose, 
 amsmath,amssymb,
 aps,
 pra,
 floatfix,
]{revtex4-1}

\usepackage{natbib,hyperref}
\hypersetup{
  colorlinks=true,urlcolor=[rgb]{0,0,1}}
\usepackage{graphicx}
\usepackage{dcolumn}
\usepackage{bm}
\usepackage{comment}
\usepackage{braket}

\begin{document}

\title{Superadiabatic driving of a three-level quantum system}

\author{M. Theisen}
\affiliation{%
    ITP, Universit\"at Heidelberg, Philosophenweg 12, 69120 Heidelberg, Germany
}

\author{F. Petiziol}
\affiliation{%
	INFN, Sezione di Milano Bicocca, Gruppo Collegato di Parma, Parco Area delle Scienze 7/a, 43124 Parma, Italy
}
\affiliation{%
	Dipartimento di Scienze Matematiche, Fisiche e Informatiche, Universit\`a di Parma,
    Parco Area delle Scienze 7/a, 43124 Parma, Italy
}

\author{S. Carretta}
\affiliation{%
	Dipartimento di Scienze Matematiche, Fisiche e Informatiche, Universit\`a di Parma,
    Parco Area delle Scienze 7/a, 43124 Parma, Italy
}

\author{P. Santini}
\affiliation{%
	Dipartimento di Scienze Matematiche, Fisiche e Informatiche, Universit\`a di Parma,
    Parco Area delle Scienze 7/a, 43124 Parma, Italy
}

\author{S. Wimberger}
\affiliation{%
    ITP, Universit\"at Heidelberg, Philosophenweg 12, 69120 Heidelberg, Germany
}
\affiliation{%
	INFN, Sezione di Milano Bicocca, Gruppo Collegato di Parma, Parco Area delle Scienze 7/a, 43124 Parma, Italy
}
\affiliation{%
	Dipartimento di Scienze Matematiche, Fisiche e Informatiche, Universit\`a di Parma,
    Parco Area delle Scienze 7/a, 43124 Parma, Italy
}

\date{\today}


\begin{abstract}
We study superadiabatic quantum control of a three-level quantum system whose energy spectrum exhibits multiple avoided crossings. In particular, we investigate the possibility of treating the full control task in terms of independent two-level Landau-Zener problems. We first show that the time profiles of the elements of the full control Hamiltonian are characterized by peaks centered around the crossing times. These peaks decay algebraically for large times. In principle, such a power-law scaling invalidates the hypothesis of perfect separability. Nonetheless, we address the problem from a pragmatic point of view by studying the fidelity obtained through separate control as a function of the intercrossing separation. This procedure may be a good approach to achieve approximate adiabatic driving of a specific instantaneous eigenstate in realistic implementations.
\end{abstract}

\maketitle


\section{Introduction}\label{sec:introduction}

The ability to control the dynamics of a quantum system is a major task for the development of quantum technologies. For this purpose, much effort has been recently invested toward the development of explicit time-dependent control protocols. 

Among the different methodologies to design control fields inducing a desired quantum evolution, the two that are attracting the most attention nowadays are optimal control theory \cite{Glaser2015} and the emergent field of so-called shortcuts to adiabaticity \cite{TORRONTEGUI2013117}. The latter, on which we will concentrate here, comprises a number of techniques whose aim is to induce perfect adiabatic evolution, that is, perfect following of the instantaneous eigenstates of the time-dependent Hamiltonian, in a finite time.

Specifically, we will focus on the shortcut known as transitionless quantum driving (TQD) \cite{berry1}, also called superadiabatic (SA) \cite{bason1} or counterdiabatic (CD) protocol \cite{rice1}. Its core idea is that, given an initial Hamiltonian $\hat H^{(0)}(t)$, it is always possible to find a correcting term $\hat H^{(1)}(t)$ which cancels nonadiabatic effects. The full Hamiltonian $\hat H(t)=\hat H^{(0)}(t)+\hat H^{(1)}(t)$ then drives the instantaneous eigenstates of $\hat H^{(0)}(t)$ exactly. 

The TQD algorithm suffers from two main weaknesses. First of all, while in principle it provides the CD control fields for quantum systems of arbitrarily many energy levels, going beyond the two-level case often becomes analytically infeasible, and must be treated numerically. Secondly, even when the CD corrections are found, they might require physical interactions which are not present in the original Hamiltonian, leading to difficulties in the experimental realizations.
The first issue will be our concern here.

Exact analytical results have been produced and tested for some specific problems, the major context being the application of the TQD protocol to improve the efficiency of ``Rapid'' adiabatic passage and stimulated Raman adiabatic passage schemes \cite{rice1, PhysRevA.94.063411, chen2, giann1, PhysRevLett.116.230503,zhou1}, in terms of fidelity, robustness, and transfer time. In addition, exact CD fields have been found for scale-invariant dynamical processes \cite{jarz1,campo1,deff1}.

In this paper, we will study the application of the SA protocol to a three-level system featuring sequences of Landau-Zener-Majorana-St\"uckelberg (LZ for brevity) crossings.
The motivation behind this choice resides in the fact that adiabatic quantum control ultimately deals with the suppression of nonadiabatic transitions, whose probability is extremely enhanced in the proximity of avoided crossings in the energy spectrum. For this reason, the LZ model \cite{zener,landau,majorana,stuckel} has been a central playground for testing control protocols \cite{bason1,malossi1}. Moreover, the LZ scenario usually yields a good local approximation to more complex spectra \cite{zenesini1,sias1,nori}.

Although for an isolated two-level LZ avoided crossing the SA algorithm provides exact analytical shapes for the control fields, the case of more-levels spectra is an open problem.

The approach which we shall adopt here differs from the typical quest for shortcuts to adiabaticity. Rather then searching alternative correcting fields which would give the exact adiabatic evolution, our main concern is to study the possibility of decomposing the TQD control Hamiltonian into the sum of separate single crossing corrections, and to test the validity of such an approximation.

After a brief review of the TQD theory and its application to the control of a single LZ anticrossing in Sec. \ref{2lev}, the system under analysis is presented in Sec. \ref{sec:hamiltonian}, together with the limiting cases where the control fields can be analytically calculated.  Section \ref{sec:controlpulses} is devoted to the discussion of the general case: the control fields are wholly determined numerically and their long-range properties are discussed according to perturbative arguments. We will see that they decay in time as power laws, preventing the definition of a natural scaling for the separability into single-crossing problems. In Sec. \ref{separability} the problem of separability of the control is addressed, by proposing possible constructions of the ``separated-control" strategy for simple experimental implementations. Numerical estimates are given for the dependence of the nonadiabatic transition probability from the time separation between the crossings, which turns out to be another power law. Such a study permits one to determine the threshold value for the time separation which is required in order to achieve a desired fidelity at the end of the protocol (i.e., a given overlap between the approximately driven and the true ground state).

We finally discuss, in Sec. \ref{espreal}, aspects concerning possible experimental realizations. We first test the robustness of our results against the introduction of asymmetries in the system under investigation, which can result from experimental constraints. The section is then concluded by proposing and analyzing possible experimental setups where the model can be realized and tested.


\section{Transitionless quantum driving}
\label{2lev}

In this section, the theory of TQD is briefly reviewed \cite{berry1,rice1}.

\subsection{General TQD theory}
Let $\hat H^{(0)}(t)$ be the $n$-level system's Hamiltonian with instantaneous eigenvalues $E_n(t)$ and eigenstates $ \ket{\psi_n(t)}$ defined by
\begin{equation}
	\hat H^{(0)}(t) \ket{\psi_n(t)} = E_n(t) \ket{\psi_n(t)} .
\end{equation}
The time evolution of a general state $\ket{\Psi(t)}$ is given by the Schr\"odinger equation
\begin{equation}
	i \hbar \frac{\partial \ket{ \Psi(t)}}{\partial t} = \hat H^{(0)}(t) \ket{\Psi(t)}.
\end{equation}

Nonadiabatic effects arising from the free evolution of $\hat H^{(0)}(t)$ can be compensated for by a control Hamiltonian $\hat H^{(1)}(t)$, such that the Hamiltonian	$\hat H(t) = \hat H^{(0)}(t) + \hat H^{(1)}(t)$
drives exactly the states
\begin{equation} \label{eq:adstates}
\exp \left[ {-\frac{i}{\hbar}\int_{t_0}^t ds E_n(s) -\int_{t_0}^t ds \braket{\psi_n(s) \lvert \partial_s \psi_n(s)}} \right] \ket{\psi_n(t)}.
\end{equation}
These are the states which would be driven in the adiabatic approximation, i.e., the exact instantaneous eigenstates of $\hat H^{(0)}(t)$ multiplied by the dynamical and geometrical phase factors.

The superadiabatic correction, with time dependencies implicitly understood in all terms, is of the form \cite{berry1}
\begin{equation}\label{eq:H1}
	\hat H^{(1)}(t) = i\hbar \sum_{m\ne n} \sum_n \frac{\ket{\psi_m} \bra{\psi_m} \partial_t \hat H^{(0)} \ket{\psi_n}\bra{\psi_n}}{E_n-E_m}.
\end{equation}
It thus depends directly on the instantaneous eigenstates. We will often refer to the off-diagonal matrix elements of $\hat H^{(1)}(t)$ as control functions, or simply controls. Note that $\hat H^{(1)}(t)$ has only off-diagonal elements in the adiabatic basis (i.e., on the basis of instantaneous eigenvectors).

The control Hamiltonian $\hat H^{(1)}(t)$ can also be rewritten by means of a matrix $\hat U(t)$ such that $\hat U(t) \hat H^{(0)}(t) \hat U^{\dagger}(t)$ is diagonal, i.e., whose rows are instantaneous eigenvectors of $\hat H^{(0)}(t)$, as \cite{rice1}
\begin{equation} \label{eq:contrUU}
\hat H^{(1)}(t) = i \hbar \frac{\partial \hat U^{\dagger}(t)}{\partial t} \hat U(t).
\end{equation}
Depending on the choice of $\hat U(t)$, the driven instantaneous eigenstates can acquire different phase factors with respect to Eq. \eqref{eq:adstates}.

Let us stress that the superadiabatic control procedure drives each eigenstate exactly, i.e., the control Hamiltonian is always the same no matter in which initial instantaneous eigenstate the system starts.


\subsection{Two-level avoided crossings} \label{sec:2levac}

The Hamiltonian for a general two-level system can be written in the real, symmetric and traceless form

\begin{align}
\hat H^{(0)}(t) & = \frac{\omega(t)}{2} \hat \sigma_z + \frac{\Delta(t)}{2} \hat \sigma_x \nonumber \\
& = \frac{1}{2} \begin{pmatrix}
\omega(t) & \Delta(t) \\
\Delta(t) & -\omega(t)
\end{pmatrix},
\end{align}
in the (time-independent) eigenbasis $\{ \ket{0}, \ket{1} \}$ of $\hat \sigma_z$, which is called the \emph{diabatic} basis. The instantaneous eigenvalues are
\begin{equation}
\pm \frac{1}{2} \sqrt{\omega(t)^2+\Delta(t)^2}.
\end{equation}
The instantaneous eigenvectors can be conveniently written by means of a trigonometric parametrization
\begin{equation}\label{eigv2}
\ket{\psi_0(t)} = \begin{bmatrix}
-\sin \frac{\theta(t)}{2} \\ \cos \frac{\theta(t)}{2}
\end{bmatrix}; 
\qquad
\ket{\psi_1(t)} = \begin{bmatrix}
\cos \frac{\theta(t)}{2} \\ \sin \frac{\theta(t)}{2}
\end{bmatrix},
\end{equation}
involving an angle defined by
\begin{equation}
\tan \theta(t) = \frac{\Delta(t)}{\omega(t)} \qquad [0 \le \theta(t) \le \pi].
\end{equation}
The LZ model is characterized by a linear sweep $\omega(t)=\alpha t$ and a constant coupling $\partial_t \Delta = 0$ \cite{zener}. The coupling induces nonadiabatic transitions in the vicinity of the avoided crossing at $t=0$, where the energy levels reach the minimum distance $\Delta$. Note that $\{\ket{\psi_0(t)},\ket{\psi_1(t)}\} \overset{t\to-\infty}{\longrightarrow } \{-\ket{0},\ket{1} \}$ and $\{\ket{\psi_0(t)},\ket{\psi_1(t)}\}\overset{t\to\infty}{\longrightarrow } \{\ket{1},\ket{0} \}$, i.e., adiabatic states connect different, orthogonal, diabatic states before and after the crossing.

In the uncontrolled case, if the system starts in the ground state $\ket{\psi_0(t)}$ at long past times, the tunneling probability exhibits an oscillatory behavior after the crossing which decays toward an asymptotic value. This value is accurately approximated by the well-known LZ formula \cite{zener}:
\begin{equation}\label{eq:P_tunnel}
P_{\text{LZ}} = \lvert\braket{\Psi(t) \lvert \psi_1(t)} \rvert^2= \exp \left( - \pi \frac{\Delta^2}{2 \hbar \alpha}\right).
\end{equation}
This expression is only valid when the system starts exactly in an eigenstate at $t\to-\infty$, while it does not hold for superposition input states.

Interestingly, the damping of the oscillations of the nonadiabatic transition probability in the diabatic basis, $\lvert\braket{\Psi(t) \lvert 1}\rvert^2$, follows a $t^{-1}$ power law \cite{limberry,vitanovgarraway}. Such a behavior can be recognized from the asymptotic properties of parabolic cylinder functions \cite{gradstheyn}, which constitute the exact solution of the Schr\"odinger equation for the coefficients in the diabatic basis, $\braket{\Psi(t) \lvert 0}$ and $\braket{\Psi(t) \lvert 1}$. This long transient makes it difficult to define a LZ transition time \cite{mullen,vitanov1,tayebirad1,zenesini1}, and suggests that consecutive crossings can never be considered as completely independent events. 

When multiple crossings between the same two levels occur, the asymptotic probability is in general not the product of LZ single-crossing probabilities, as given in Eq. \eqref{eq:P_tunnel}. The relative phase accumulated during the evolution among different crossings, for instance, gives rise to interference phenomena showing up as St\"uckelberg oscillations  \cite {stuckel,nori}. Such phenomena are particularly relevant for systems subject to periodic potentials \cite{nori,PhysRevLett.105.215301,PhysRevA.82.065601}.

The control Hamiltonian $\hat H^{(1)}(t)$ for the general two-level system can be exactly calculated from \eqref{eq:H1}, using the expressions \eqref{eigv2} of the eigenvectors:
\begin{equation} \label{eq:2levcontrol}
\hat H^{(1)}(t) = \frac{1}{2} \frac{\partial \theta (t)}{\partial t} \hat \sigma_y,
\end{equation}
where, in terms of the parameters of the Hamiltonian,
\begin{equation}
\frac{\partial \theta(t)}{\partial t} = \frac{\dot \Delta(t) \omega(t) - \Delta(t) \dot \omega(t)}{\omega(t)^2+\Delta(t)^2} ,
\end{equation}
where the dot denotes time derivative.
As was pointed out in \cite{rice2}, $\partial_t \theta(t)$ has typically a Lorentzian shape in the vicinity of an avoided crossing. This is indeed verified in the LZ model:
\begin{equation} \label{eq:LZcontrol}
\frac{\partial \theta_{\text{LZ}}(t)}{\partial t} = - \frac{\frac{\Delta}{\alpha}}{t^2+\left(\frac{\Delta}{\alpha}\right)^2}. 
\end{equation}
Time integration of Eq. \eqref{eq:LZcontrol} shows that 
\begin{equation}\int_{-\infty}^{\infty} \partial_t \theta_{\text{LZ}}(t) dt=\pi.\end{equation}
Therefore $\partial_t \theta_{\text{LZ}}(t) $ represents a $\pi$ pulse.


\section{Three-level system}\label{sec:hamiltonian}
\begin{figure}
    \includegraphics[width=\linewidth]{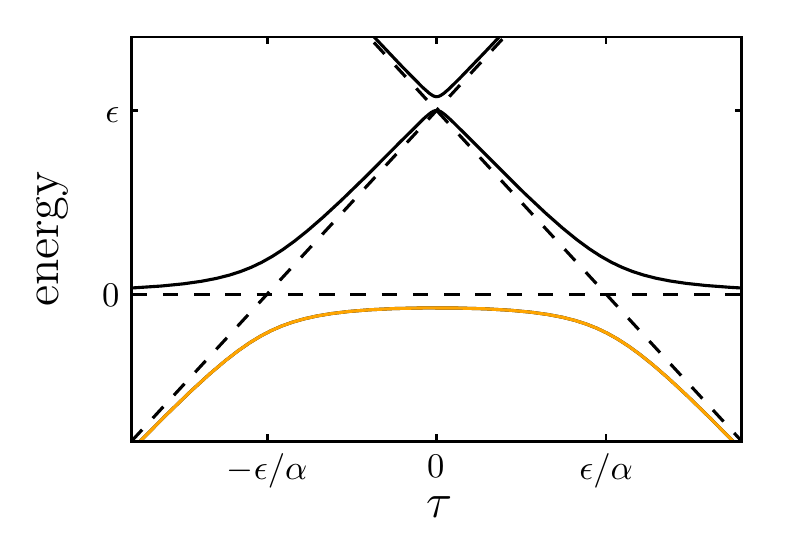}
    \caption{
    	Time evolution of the energy spectrum for $\omega(\tau)=\alpha \tau$. Dashed and solid lines represent diabatic and adiabatic eigenstates, respectively. The temporal evolution of the ground state subject to superadiabatic control is given by the orange line.
    	}
	\label{fig:spec}
\end{figure}
In this paper, we study the specific three-level Hamiltonian

\begin{equation}\label{eq:hamiltonian}
	\hat H^{(0)}(\tau) = 
    \begin{pmatrix}
    	\epsilon + \omega(\tau) &\frac{\Delta}{\sqrt{2}} & 0 \\
    	\frac{\Delta}{\sqrt{2}} & 0 & \frac{\Delta}{\sqrt{2}} \\
        0 & \frac{\Delta}{\sqrt{2}} & \epsilon - \omega(\tau)
    \end{pmatrix},
\end{equation}
with a linear sweep function $ \omega(\tau)=\alpha \tau$ implementing the LZ dynamics.
The Hamiltonian and its parameters are considered to be dimensionless, therefore scaled with respect to some characteristic energy of the system $E^{(c)}$ which defines our unit of energy. Time is accordingly rescaled in the dimensionless form $\tau=E^{(c)} t/\hbar$. In terms of spin-1 operators, $\hat H^{(0)}(\tau) = \epsilon \hat S_z^2 + \omega(\tau) \hat S_z + \Delta \hat S_x$, and the dimensionless Schr\"odinger equation reads $i \partial_{\tau} \ket{\psi} = \hat H^{(0)}(\tau) \ket{\psi} $.
Furthermore, let $\{\ket{1},\ket{2},\ket{3}\}$ be the diabatic basis, i.e., the (time-independent) eigenvectors of $\hat H^{(0)}(\tau)$ for $\Delta=0$.

Similar three-level Hamiltonians have been studied in \cite{poggi1}, from a time-optimal control perspective, and in \cite{kiselev1}, where the behavior of the transition probability in the absence of controls is inspected.

Figure \ref{fig:spec} displays the time evolution of the instantaneous energy spectrum, where dashed and solid lines are used for the diabatic and adiabatic levels, respectively. Thus, the Hamiltonian describes a system where two diabatic energy levels are indirectly coupled through an intermediate one ($\{\hat H^{(0)}\}_{12}=\{\hat H^{(0)}\}_{23}=\Delta$, but $\{\hat H^{(0)}\}_{13}=0$). Such ``indirect" interaction induces a narrow avoided crossing at $\tau=0$ between the two larger, ``direct'' ones at $\omega(\tau)=\pm \epsilon$. The parameter $\epsilon$, together with the sweep rate $\alpha$, characterizes therefore the separation of crossings.
Our aim is to quantify the effect of the indirect coupling on the system's dynamics. More precisely, we study how its presence affects the control Hamiltonian, and to which extent it can be neglected when one is interested in guaranteeing adiabatic evolution only for the ground state. 
As a first step, let us discuss the analytically accessible limits for the control Hamiltonian.
 
The straightforward calculation of the control Hamiltonian $\hat H^{(1)}(\tau)$ requires knowledge of the instantaneous eigenstates of the system, as evident by Eq. \eqref{eq:H1}.
The problem of calculating them exactly for systems with more than two levels constitutes the main limit of the superadiabatic approach.

For the three-level case, the eigenstates are in general accessible through Cardano's \emph{casus irreducibilis}. However, the resulting exact expressions for the eigenvalues are typically difficult to handle, even via symbolic computing packages, especially due to their complex-valued representation. For this reason, we will not take this route, but rather solve limiting cases analytically and attack the general case numerically.

Firstly, let us regard the case of no separation $\epsilon = 0$, where all diabatic levels cross in a single point at $\tau=0$ and the three anticrossing coalesce into a single one. 

The eigenvalues are (labeling increases for increasing energy)

\begin{equation}
\lambda_1^{\epsilon=0}(\tau) = 0 ; \, \, \, \lambda_{2,0}^{\epsilon=0}(\tau) = \pm \sqrt{\omega(\tau)^2+ \Delta^2},
\end{equation}
and the instantaneous eigenvectors are the rows of
\begin{equation}\label{eq:eigeps0}
\hat U^{\epsilon=0}(\tau) = \begin{pmatrix}
\sin^2 \frac{\phi}{2} & -\frac{1}{\sqrt{2}} \sin \phi  & \cos^2 \frac{\phi}{2} \\
-\frac{1}{\sqrt{2}} \sin{ \phi}  & 2 \cos^2 \frac{\phi}{2} -1 & \frac{1}{\sqrt{2}} \sin \phi \\
\cos^2 \frac{\phi}{2} & \frac{1}{\sqrt{2}}\sin  \phi & \sin^2 \frac{\phi}{2}
\end{pmatrix},
\end{equation}

where the angle $\phi(\tau)$ is defined by $\tan\phi(\tau) = \Delta/\omega(\tau)$. By Eqs. \eqref{eq:eigeps0} and \eqref{eq:contrUU} the CD Hamiltonian in the diabatic basis takes the simple form
\begin{equation} \label{eq:H1sym}
	\hat H^{(1)}(\tau; \epsilon=0) = \frac{\partial \phi(\tau)}{\partial \tau} \hat S_y.
\end{equation}

Therefore $\partial_{\tau} \phi(\tau)$ is a Lorentzian pulse of the form given in Eq. \eqref{eq:LZcontrol} for linear energy sweeps, as in the two-level LZ case. As for $\partial_{\tau} \theta_{\text{LZ}}(\tau)$ of Eq. \eqref{eq:2levcontrol}, it holds that $\int_{-\infty}^{\infty}\partial_{\tau} \phi(\tau) d\tau = \pi$. 

The second solvable limit is the instant $\tau = 0$, for a generic $\epsilon\ne 0$.
The eigenvalues in this case are
\begin{equation} \label{eq:eigt0}
\lambda_1(0) = \epsilon; \quad \lambda_{2,0}(0)= \frac{\epsilon}{2} \pm \frac{1}{2}\sqrt{\epsilon^2 + 4 \Delta^2},
\end{equation}

leading to the control Hamiltonian
\begin{equation}\label{eq:H_CD1}
    \hat H^{(1)}(0) = \frac{i}{\sqrt{2}} \left.\frac{\partial \phi(\tau)}{\partial \tau} \right|_{\tau=0}
    \begin{pmatrix}
        0 & -1  & -\sqrt{2} \epsilon/\Delta \\
        1 & 0 & -1 \\
        \sqrt{2} \epsilon/\Delta & 1 & 0
    \end{pmatrix}.
\end{equation}
In comparison with $\hat H^{(1)}(\tau;\epsilon=0)$, Eq. \eqref{eq:H1sym}, we observe that $\hat H^{(1)}(0)$ introduces an additional off-diagonal control element, whose maximal intensity scales linearly with $\epsilon$.

Note that, from Eq.~\eqref{eq:eigt0}, we can extrapolate the minimum distance, i.e., the distance at resonance, between the two higher energy levels to be $\tilde \Delta \equiv \lambda_2(0) - \lambda_1(0) = \frac{\epsilon}{2} \left(\sqrt{4 \Delta^2/\epsilon^2+1}-1 \right)$. This suggests that the corresponding anti-crossing may be locally approximated by a two-level LZ with coupling $\tilde \Delta/2 $ (see Sec. \ref{sec:2levac}). In particular, for small ratios $\Delta/\epsilon$ the coupling to first order is given by $\tilde \Delta/2  \simeq  \Delta^2/2 \epsilon$. Although this two-level approximation well describes the local behavior of the energy levels, we will see that it is not the case for the associated control function. 


\section{Control Pulses}\label{sec:controlpulses}

\begin{figure}
    \includegraphics[width=\linewidth]{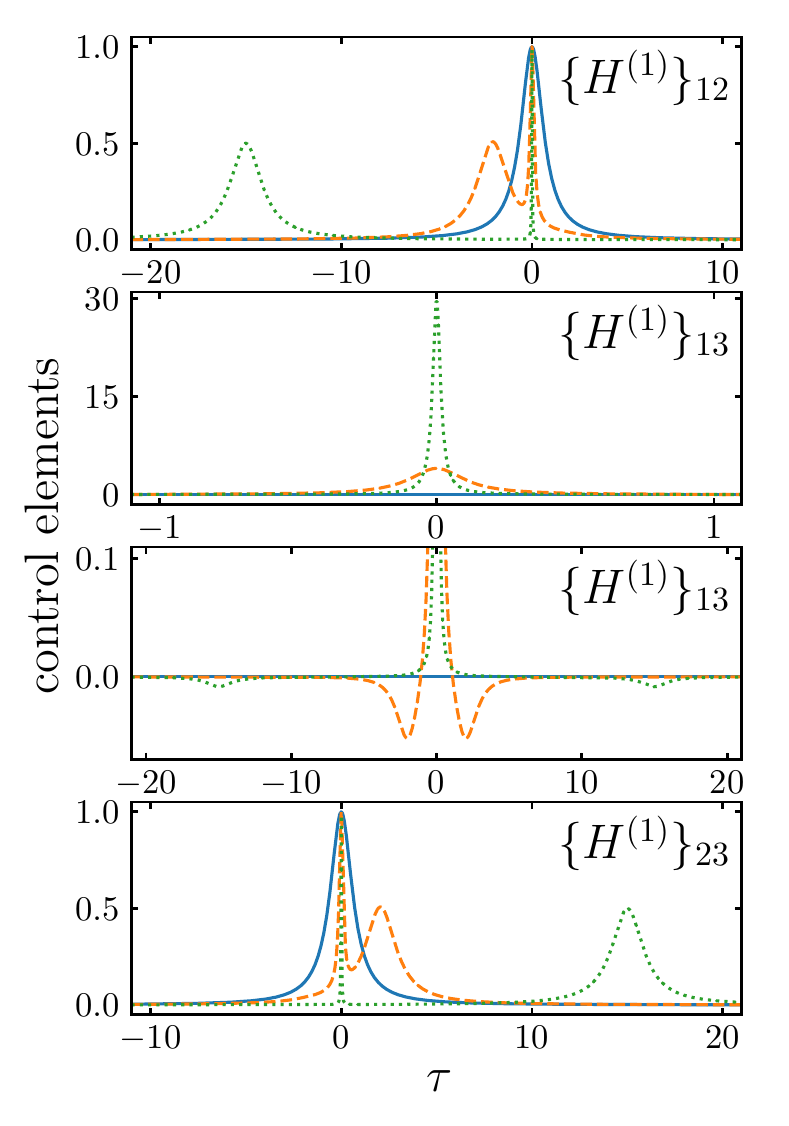}
    \caption{
    	Shape of the control functions, i.e., of the matrix elements of the control Hamiltonian $H^{(1)}(t)$ for three different separations of the avoided crossings.
        Solid, dashed, dotted lines correspond to $\epsilon = 0, 2,$ and $15$, respectively. Parameters are
        $\alpha=1$, $\Delta=0.5$.
   	}
	\label{fig:control_elements}
\end{figure}

In order to study the problem for any (dimensionless) time $\tau$ and separation $\epsilon$ of crossings, we compute numerically the full control Hamiltonian $\hat H^{(1)}(\tau)$ according to Eq.~\eqref{eq:H1}. Note that $\hat H^{(1)}(\tau)$ drives any of the three eigenstates exactly at the same time, although we are mainly concerned with the ground state, which undergoes two avoided crossings.
In Fig.~\ref{fig:control_elements} the three independent control functions constituting $\hat H^{(1)}(\tau)$ are shown for different $\epsilon$. In the analytically solvable case $\epsilon=0$ (solid lines) the peaks are Lorentzian functions centered around $\tau=0$. For small (dashes lines) and large (dotted lines) $\epsilon$, the peaks of $\{\hat H^{(1)}\}_{12}$ and $\{\hat H^{(1)}\}_{23}$ shift with respect to the origin. Their center, i.e. the position of the anti-crossings, is given by $\tau=\pm \epsilon/\alpha$ and their maximum reduces to half the value they have for $\epsilon=0$. Note that a sharp peak of constant intensity located at the origin is still present. 

The interesting element $\{\hat H^{(1)}\}_{13}$, appearing only for $\epsilon\neq0$, also features a sharp peak at the origin which has maximal value proportional to $\epsilon$, as can be seen from \eqref{eq:H_CD1}. However, the sign of the other peaks is negative and their intensity is small compared to all other peaks. This is due to the fact that, when the two larger avoided crossings occur, the levels involved are also indirectly coupled through the third level. This indirect interaction contributes to the formation of the avoided crossing, but as a second-order effect in comparison with the direct interaction. 
In order to justify the latter assertion, we perturbatively study the system in the limit of small $\Delta$, by treating $\hat V = \Delta \hat S_x$ as a perturbation. At the lowest orders in $\Delta$, stationary perturbation theory gives the following profiles for the control functions:

\begin{align}
	\{\hat H^{(1)}\}_{12} &= i\frac{\dot\omega}{(\epsilon + \omega)^2} \frac{\Delta}{\sqrt{2}}
    	+ O(\Delta^3) ,\\
	\{\hat H^{(1)}\}_{23} &= i\frac{\dot\omega}{(\epsilon - \omega)^2} \frac{\Delta}{\sqrt{2}}
    	+ O(\Delta^3) ,\\
	\{\hat H^{(1)}\}_{13} &= i\frac{\dot\omega \epsilon \left(\epsilon^2-5 \omega^2\right)}
    	{4 \omega^2\left(\omega^2-\epsilon^2 \right)^2} \Delta^2
    	+ O(\Delta^4).
\end{align}

While the leading order for the control elements related to the ``direct" crossings is $\sim\Delta$, the one for the control of the ``indirect" crossing is $\sim\Delta^2$. This witnesses the fact that the narrower crossing at $\tau=0$ is the result of a nonadiabatic coupling acting at second order in $\Delta$. Perturbation theory, as is often the case, allows one to reinterpret the ``indirect" interaction in terms of an effective direct coupling, which shows up at the second order in $\Delta$. This also gives a clearer picture of the hierarchy of interactions, which is reflected by the relative intensity of the peaks in the control functions, as previously mentioned above. 

Let us now focus on the long-range decay of the control functions. Such a study, as can be seen from the definition of $\hat H^{(1)}$ in Eq. \eqref{eq:H1}, requires knowledge of the dominant long-time behavior of the instantaneous energies and eigenvectors of $\hat H^{(0)}(t)$. In order to obtain it, we develop a slightly different perturbative approach: we recast the Hamiltonian into a form which allows us to treat $\lambda = \tau^{-1}$ as a perturbative parameter and subsequently apply standard perturbation theory again. 
This is done by considering the Hamiltonian $\hat H(\tau)/\tau$, separated into the two terms  $\alpha \hat S_z + \lambda (\epsilon \hat S_z^2+\Delta \hat S_y) $. Perturbation theory can then be applied by taking $V = \lambda (\epsilon \hat S_z^2+\Delta \hat S_x^2)$ as a perturbation, and provides a first-order expansion of the instantaneous energies and eigenvectors. The latter are the same as for the original $\hat H(\tau)$, while the instantaneous energies can be reobtained by multiplication by $\tau$ of the perturbative energy expansion. We then use Eq. \eqref{eq:H1} to compute the $\lambda$-leading terms of the matrix elements of $\hat H^{(1)}$. 
Let us remark that, as the intensity peaks are located at the crossing centers, where the energy gaps reach their minimum values, the use of perturbation theory is fully justified only for studying the behavior of the tails, which is exactly our purpose. 

The off-diagonal matrix elements of the control Hamiltonian, up to second order, behave like
\begin{equation} \label{eq:timecontr}
	\hat H^{(1)}_\text{pert} \propto i
	\begin{pmatrix}
    	0 & \Delta\tau^{-2} & -\Delta^2\tau^{-4}
    \\
        - \Delta\tau^{-2} & 0 & \Delta\tau^{-2}
    \\
        \Delta^2\tau^{-4} & -\Delta\tau^{-2} & 0
    \end{pmatrix}.
\end{equation}

\begin{figure}
    \includegraphics[width=\linewidth]{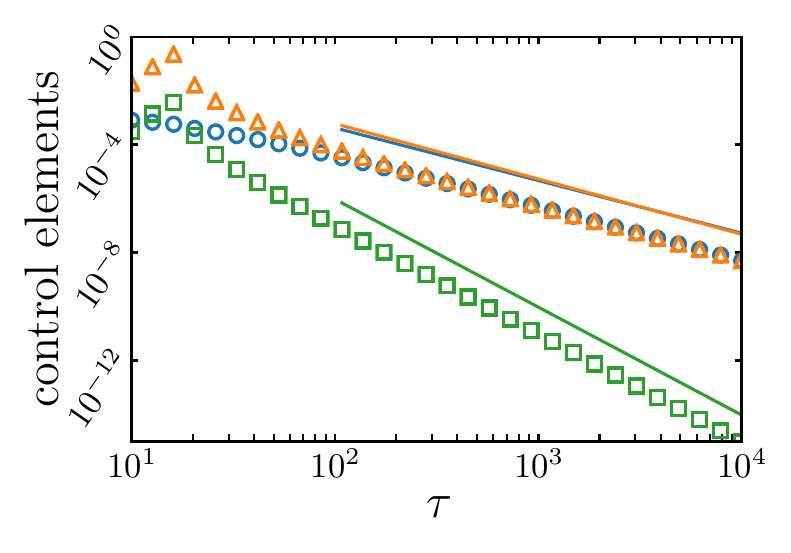}
    \caption{
    	Asymptotic (large $\tau$) behavior of control elements for $\epsilon=15$. Circles, triangles, and squares are used for the elements $\{\hat H^{(1)}\}_{12}$, $\{\hat H^{(1)}\}_{23}$, and $\{\hat H^{(1)}\}_{13}$, respectively. The power-law fits are shifted for better visibility. The respective fit coefficients are $-2$, $-2$, and $-4$. Parameters are $\alpha=1$, $\Delta=0.5$.
   	}
	\label{fig:control_elements2}
\end{figure}

We now see that all elements decay algebraically in the asymptotic limit, and this is indeed verified by our numerics, as shown in Fig.~\ref{fig:control_elements2}. This behavior is a negative point concerning the separability of crossings: the power-law scalings prevent the possibility of obtaining perfectly separable controls, since no definite decay scale can be defined. Note that, although the ``direct" coupling elements $\{\hat H^{(1)}\}_{12}$ and $\{\hat H^{(1)}\}_{23}$ follow the familiar Lorentzian $\tau^{-2}$ dependence, the ``indirect" one $\{\hat H^{(1)}\}_{13}$ decays faster, as $\tau^{-4}$. Actually, terms of order $\tau^{-3}$ also appear in the perturbative calculation for the ``indirect" crossing, but they cancel each other exactly due to the temporal symmetry of the problem. The exact cancellation no longer happens when such symmetry is broken, and this feature will be discussed in Sec.~\ref{sec:asymmetries}.


\section{separability}\label{separability}
We have seen that, due to the power-law scaling of the control functions, the global control problem cannot be fully decomposed into the sum of single-crossing corrections. Nevertheless, we investigate whether efficient approximate control can still be achieved by separately correcting each single anticrossing.

The general strategy is that each time the system undergoes an avoided crossing, a superadiabatic two-level control pulse is applied in order to drive the system transitionlessly. The general shape of such a pulse is given in Sec. \eqref{sec:2levac}, but the specific Hamiltonian implementing it is constrained by the experimental resources. In particular, it depends on which single matrix elements of the Hamiltonian can be externally controlled. The only essential requirement is that the pulse acts in an effective manner only on the two levels involved in the crossing, without activating undesired tunneling effects with or among other levels. Here we discuss two possible constructions of the ``separated-control" Hamiltonian. 
 
We focus on driving of the instantaneous ground state of our system along the two sequential LZ crossings it goes through, ignoring the third anticrossing between the other states. This is motivated by the fact that many quantum control problems require adiabatic following of exactly one of the instantaneous eigenstates. This is true especially in view of situations where the number of anticrossings in a multilevel spectrum is very large, although the evolution of only one instantaneous eigenstate is of interest.
The efficiency of the method is quantified by how close the state driven by the separated control field is to the exact instantaneous ground state at the end of the protocol.
In particular, we will study numerically the probability of nonadiabatic transitions. This is defined as 
\begin{equation}\label{nonadP} \mathcal P = 1- \braket{\Psi \lvert \psi_0},\end{equation}
i.e., as the deviation from one of the overlap between the driven state $\ket{\Psi(\tau)}$ and the true instantaneous ground state $\ket{\psi_0(\tau)}$.

The first construction we consider is the following. We initially design the control Hamiltonians which would drive the system transitionlessly if only one of the two anticrossings were present; that is, which drive exactly the instantaneous eigenstates of Eq. \eqref{eq:hamiltonian} when $\{\hat H^{(0)}\}_{23}=\{\hat H^{(0)}\}_{32}=0$ or $\{\hat H^{(0)}\}_{12}=\{\hat H^{(0)}\}_{21}=0$, respectively. We then construct the full control field by summing the separate ones. 

\begin{figure}
    \includegraphics[width=\linewidth]{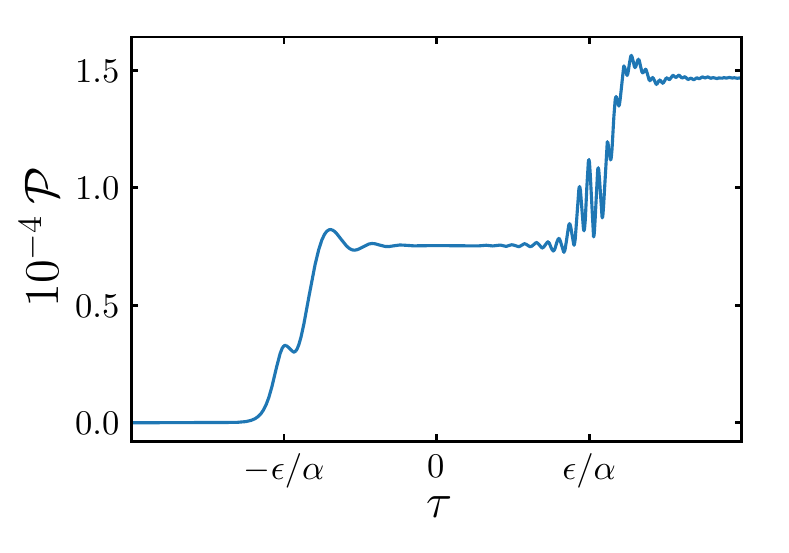}
    \caption{
		Time evolution of the probability of nonadiabaticity $\mathcal P$, defined in Eq. \eqref{nonadP}, for $\alpha=1$, $\Delta=0.5$, $\epsilon=7$. The fact that the separate control is not perfectly driving the adiabatic states shows up in terms of residual (not canceled by the control) ``jumps" of $\mathcal P$ in the vicinity of the anticrossings.}
	\label{fig:P_time}
\end{figure}

In order to define the separate control Hamiltonians, let us introduce the matrices $\hat U_L(\tau)$ and $\hat U_R(\tau)$ which diagonalize, respectively, the upper-left and lower-right two-by-two submatrices of $\hat H^0(\tau)$, corresponding to the left $(L)$ and right $(R)$ anticrossings. From Eq. \eqref{eigv2}, they are defined by
\begin{subequations}
\begin{equation}
\hat U_{L}(\tau) = \begin{pmatrix}
-\sin \frac{\theta_{L}(\tau)}{2} & \cos \frac{\theta_{L}(\tau)}{2} & 0 \\
\cos \frac{\theta_L(\tau)}{2} & \sin \frac{\theta_L(\tau)}{2} & 0 \\
0 &  0 & 1 
\end{pmatrix},
\end{equation}
\begin{equation}\label{eq:U2sep}
\hat U_R(\tau) = \begin{pmatrix}
1 & 0 & 0 \\
0 & -\sin \frac{\theta_R(\tau)}{2} & \cos \frac{\theta_R(\tau)}{2} \\
0 & \cos \frac{\theta_R(\tau)}{2} & \sin \frac{\theta_R(\tau)}{2}  \\
\end{pmatrix} ,
\end{equation}
\end{subequations}
with the angles being defined by
\begin{equation}
\tan \theta_L(\tau) = \frac{\sqrt{2} \Delta}{\omega(\tau)+\epsilon} ; \qquad \tan \theta_R(\tau) = \frac{\sqrt{2} \Delta}{\omega(\tau)-\epsilon}.
\end{equation}

The ``separated-control" Hamiltonian $\hat H^{(1)}_\text{sep}(\tau)$ is then constructed by summing the two control Hamiltonians correcting the $(L)$ and $(R)$ anticrossings, respectively. Recalling the definition of the control Hamiltonian for the single crossing from Eq. \eqref{eq:contrUU}, one obtains
\begin{align}
\hat H^{(1)}_\text{sep}(\tau) & = i \frac{\partial \hat U_L^{\dagger}}{\partial \tau} \hat U_L+ i \frac{\partial \hat U_R^{\dagger}}{\partial \tau} \hat U_R \label{eq:sepcontr}\\
& = \frac{i }{2} \begin{pmatrix}
0 & -\partial_{\tau} \theta_L(\tau) & 0 \\
\partial_{\tau} \theta_L(\tau) & 0 & -\partial_{\tau } \theta_R(\tau) \\
0 & \partial_{\tau} \theta_R(\tau) & 0
\end{pmatrix}. \label{eq:sepcontrol}
\end{align}
This construction is practically useful when one has independent control over the $[(12)-(21)]$ and $[(23)-(32)]$ control elements.
It must be stressed that this procedure approximately drives the population of the instantaneous ground state to be close to one but does not provide a precise control on phase factors, in contrast to the exact recipe of Eq. \eqref{eq:H1}. This is the case even for scenarios in which the anticrossings are largely separated.

In a possible different construction, the separated control can be realized through a single field modulated in time with a profile $\partial_{\tau} \theta_{\text{sep}}$ which is the sum of the profiles of the single corrections. That is, $\hat H_{\text{sep}}^{(1)}= \partial_{\tau} \theta_{\text{sep}} \hat S_y/\sqrt{2}$ with
\begin{align}
\partial_{\tau} \theta_{\text{sep}}&  = \partial_{\tau} \theta_L(\tau) + \partial_{\tau} \theta_R(\tau) \nonumber \\
& = - \frac{\sqrt{2} \alpha \Delta}{(\epsilon+\alpha \tau)^2 + 2 \Delta^2} - \frac{\sqrt{2} \alpha \Delta}{(\epsilon-\alpha \tau)^2 + 2 \Delta^2}. \label{sepprofileBy}
\end{align}
As we will also discuss in Sec. \ref{sec:exper}, this construction, for instance, is useful when the model describes a magnetic driving of a pure spin-1 system. In such a case, it is not obvious how to control independently the matrix entries [$(12)-(21)$] vs [$(23)-(32)$], while it is natural to control all of them at once by means of a single component of the magnetic field. What is important in this situation is that the control pulse, while correcting the $(L)$ crossing for example, is not intense enough to activate tunneling also toward the third level.

\begin{figure}
    \includegraphics[width=\linewidth]{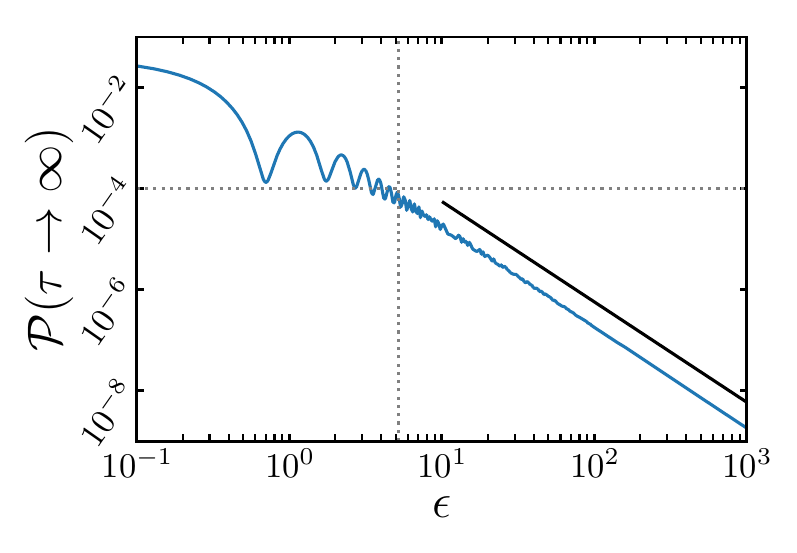}
    \caption{
		Asymptotic (large $\tau$) probability of nonadiabaticity $\mathcal P$ as a function of the intercrossing separation $\epsilon$ for fixed parameters $\alpha=\Delta=1$. The power-law fit is shifted for better visibility with the fit coefficient being $-2$. The dashed lines highlight the threshold value $\epsilon\approx 5$ for $\mathcal{P}\sim 10^{-4}$.
    	}
	\label{fig:P_na}
\end{figure}
In order to inspect numerically the efficiency of the separated control (for both constructions), we prepare the system in the ground state and propagate it in time using $\hat H(\tau) = \hat H^{(0)} + \hat H^{(1)}_\text{sep}$. 

We then extract the asymptotic probability of nonadiabaticity $\mathcal P$, as defined in Eq. \eqref{nonadP}. The fact that the correction is not perfect implies the amplification of $\mathcal P(\tau)$ in the vicinity of the anticrossing times (see Fig. \ref{fig:P_time}). The oscillatory behavior resembles the typical LZ transient. The asymptotic value of $\mathcal P$ as a function of the intercrossing separation $\epsilon$, with $\alpha$ fixed and for the first construction, Eq. \eqref{eq:sepcontrol}, is plotted in Fig. ~\ref{fig:P_na}. As can be seen, $\mathcal P$ oscillates for small $\epsilon$. This is due to St\"uckelberg's phenomenon: firstly the driven instantaneous ground state, when approaching the right crossing, is actually a superposition of the true instantaneous eigenstates. Moreover, the presence of the central anticrossing, which has been neglected, implies that the right crossing involves the same two energy levels as the left one. Therefore, the relative phase accumulated by the corresponding instantaneous eigenstates during the evolution between the $(L)$ and $(R)$ crossings also contributes to the formation of St\"uckelberg oscillations.

For large $\epsilon$, however, the oscillations in the probability are damped as the propagated states approach the exact ones, eventually following a power law as $\mathcal P\propto\epsilon^{-2}$. The same holds for a separated control of the form given in Eq. \eqref{sepprofileBy}. Due to the power-law scaling no natural threshold for $\epsilon$ can be defined at which the two crossings can be considered fully separable. However, given a desired fidelity one can extract the corresponding critical $\epsilon$, or vice versa, from Fig.~\ref{fig:P_na}. As an example, in the case of the first construction of the separated-control Hamiltonian, for $\mathcal P\sim 10^{-4}$ we get $\epsilon \approx 5$.


\section{Toward experimental realizations}\label{sec:asymmetries} \label{espreal}

In this section, we first discuss the robustness of our results with respect to the introduction of little asymmetries in the time-dependent spectrum. This is an important point for experimental realizations, as the highly symmetric spectral configuration previously studied (Fig. \ref{fig:spec}) could be hard to reproduce in practice. Possible experimental setups will be subsequently addressed where the model investigated here could be implemented and tested. 

\subsection{Asymmetric spectra}

We study two possible sources of asymmetry. The first one is to consider that the diabatic couplings inducing the $(L)$ and $(R)$ anticrossings are not exactly the same (both of strength $\Delta$), but they rather differ by a small quantity $\delta \Delta$. The second concerns instead the structure of the spectrum: we will deform the ``isosceles" triangular configuration by tilting one of the diabatic levels. The temporal $\tau\leftrightarrow-\tau$ symmetry is in fact broken by changing the slope of one of the tilted diabatic levels. This is done through the generalization $-\alpha\to \beta$.

\begin{figure}
    \includegraphics[width=\linewidth]{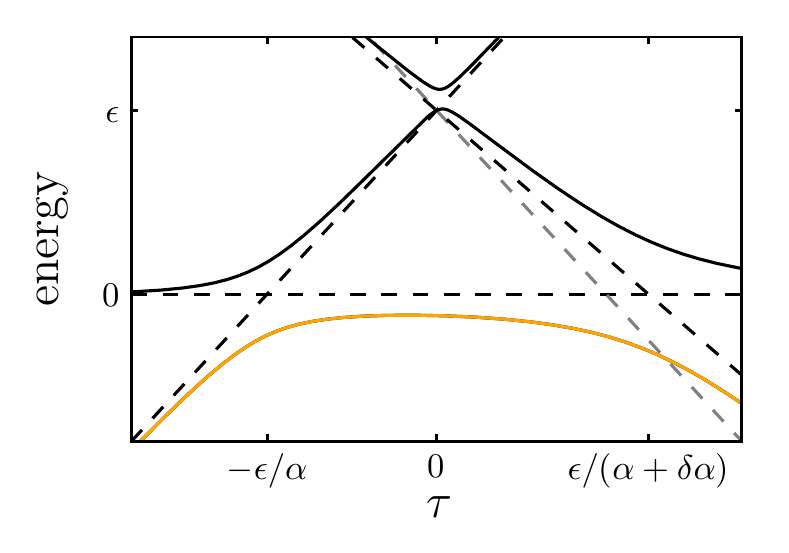}
    \caption{
    	Time evolution of the energy spectrum for the asymmetric problem for $\delta\Delta = 0.5\Delta$ and $\delta\alpha = -0.2\alpha$. Black dashed and solid lines represent diabatic levels and adiabatic levels, respectively.  The gray dashed line marks the symmetric diabatic level for $\delta\alpha = 0$. Large parameters have been used for the purpose of visibility in the plot.    	}
	\label{fig:specasym}
\end{figure}

Let us accordingly change the initial Hamiltonian $\hat H^{(0)}(\tau)$ of Eq. \eqref{eq:hamiltonian} to be:
\begin{equation}
\begin{pmatrix}
\epsilon + \alpha \tau & \Delta/\sqrt{2} & 0 \\
\Delta/\sqrt{2} & 0 &\frac{1}{\sqrt{2}}( \Delta + \delta \Delta ) \\
0 & \frac{1}{\sqrt{2}} (\Delta+\delta \Delta) & \epsilon+ \beta \tau
\end{pmatrix}.
\end{equation}
The new temporal evolution of the instantaneous spectrum is depicted in Fig. \ref{fig:specasym}.

Let us first concentrate on the first kind of asymmetry by setting $\beta= -\alpha$.
Repetition of the perturbative argument developed in Sec. \ref{sec:controlpulses} gives the same long-time scalings for the control functions as in the symmetric case. Namely, a power-law behavior with $\tau^{-2}$ for the control of ``direct'' crossings and $\tau^{-4}$ for ``indirect" crossings, as in Eq. \eqref{eq:timecontr}. This is indeed verified by numerical investigation. 
Furthermore, also the $\epsilon^{-2}$ scaling of the asymptotic infidelity $\mathcal P$, defined in Eq. \eqref{nonadP}, in the case of application of separate controls, is verified. This is reasonable as a slight variation of a constant parameter of the problem does not influence the time behavior of the control functions.

\begin{figure}
    \includegraphics[width=\linewidth]{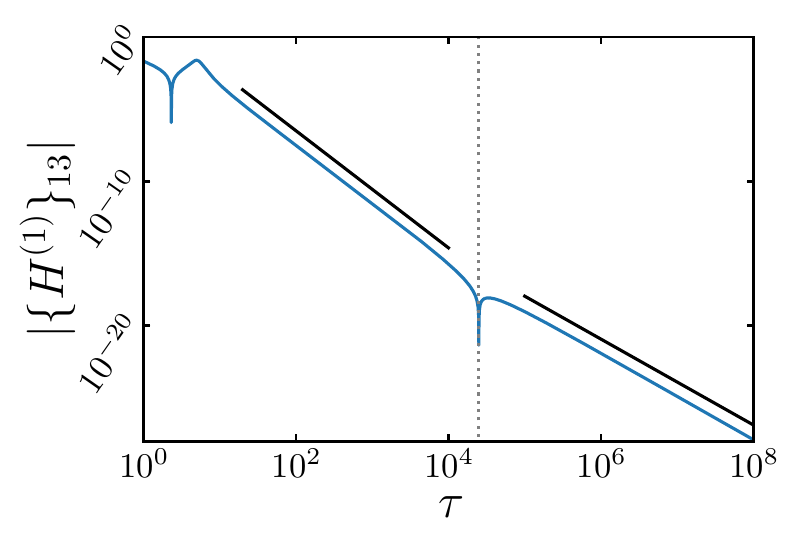}
    \caption{
    	Asymptotic (large $\tau$) behavior of the $\{\hat{H}^{(1)} \}_{13}$ control element for the asymmetric problem with $|\delta\alpha|=10^{-3}$. The other parameters are $\alpha=1$, $\Delta=0.5$, and $\epsilon=5$. Log-linear fits are shifted for greater visibility with fit coefficients $-4$ and $-3$ for the left and right line, respectively. Two regimes $\sim\tau^{-4}$ and $\sim\tau^{-3}$ are recognizable. The dotted gray line marks the transition time at approximately $\tau\simeq 5\epsilon/|\delta\alpha|$.
    	}
	\label{fig:asymcontr}
\end{figure}

Let us now consider $\delta \Delta=0$ and $\beta \ne -\alpha$. This way, we focus on the second kind of asymmetry. We then repeat once again the perturbative calculation. Although for $ \{\hat H^{(1)}\}_{12}$ and $\{\hat H^{(1)}\}_{23}$, i.e., for the $(L)$ and $(R)$ crossings, we obtain a $\tau^{-2}$ scaling of the corresponding control functions, for $\{\hat H^{(1)}\}_{13}$ we have
\begin{align}
 \{\hat{H}^{(1)}\}_{13} & = i \frac{(\alpha+\beta) \Delta^2}{2 \alpha \beta (\alpha-\beta) \tau^3} \nonumber\\
 & - i\frac{(2 \alpha^2-\alpha \beta +2 \beta^2) \epsilon \Delta^2}{2 \alpha^2 \beta^2 (\alpha-\beta) \tau^4} + O(\tau^{-5}).
\end{align}
We see thus that the asymmetry introduces a $\tau^{-3}$ term in $ \{\hat H^{(1)}\}_{13}$. Nonetheless, in order to compare its dominance with respect to the $\tau^{-4}$ term, we study both in the limit in which $\beta$ is close to $-\alpha$, let us say $\beta=- (\alpha+\delta \alpha)$ with $\delta \alpha \ll 1$. To the lowest order in $\delta \alpha$, the two competing terms have the form
\begin{equation}
C_1 \frac{\delta \alpha}{\tau^3} +  \frac{C_2}{\tau^4},
\end{equation}
with $C_1 = \Delta^2/4 \alpha^3$ and $C_2=5 \epsilon \Delta^2/4 \alpha^3$.
Note that the ${\tau^{-3}}$ term is at least proportional to $\delta \alpha$, while to the lowest order the ${\tau^{-4}}$ term is independent from $\delta \alpha$.
Therefore, depending on the $\delta \alpha$ and $\tau$ regimes, one of the two terms will be dominant (Fig. \ref{fig:asymcontr}).

The $\tau^{-4}$ term is dominant in the perturbed system only for times
\begin{equation}\tau < \frac{C_2}{C_1|\delta \alpha|}= \frac{5 \epsilon}{ |\delta \alpha|}. \end{equation}
The $\epsilon^{-2}$ dependence of the asymptotic infidelity $\mathcal P$ in the case of separate control is confirmed also for the second kind of asymmetry. This is understandable because, even if the presence of $\delta \alpha$ deforms the triangular pattern of diabatic levels, for small $\delta \alpha$ the parameter $\epsilon/\alpha$ still remains the relevant scale of separation of the avoided crossings.

\subsection{Experimental realizations}
\label{sec:exper}

The engineering of few-level systems is, in principle, possible in a number of physical situations. We briefly discuss a variety of such realizations, highlighting their main advantages and disadvantages in relation to the reproduction of the model studied in this paper.

The use of Bose-Einstein condensates \cite{becpethick,becpitaevs} offers independent control of single parameters and has the advantage that coherence time scales are typically much longer than in solid-state realizations. In such a context, the two-level SA protocol as presented in Sec. \ref{2lev} has been implemented by means of a condensate loaded into an accelerated optical lattice \cite{bason1}: two energy bands were controlled to study LZ transitions with time-dependent \textit{in situ} parameter variations. The same idea may be extended in order to control three energy bands reproducing our system of Hamiltonian \eqref{eq:hamiltonian} coupled in sequence at the respective energy gaps. Full control over three bands, however, is hard due to the fact that band gaps are very different and hence leakage toward higher bands can occur (see \cite{zenesini2008,tayebirad1,reviewWimb}).

Another possible realization with ultracold noninteracting atoms in magnetic traps was proposed already back in 1997 by Vitanov and Suominen \cite{vitanov1997}, yet in a quite different context. The implementation of the SA control protocol with this system, where individual levels correspond to internal atomic hyperfine states controlled by magnetic fields, may allow one to verify our predictions.

A strong interaction between ultracold bosons may also be exploited to implement a level structure corresponding to Fig. \ref{fig:spec} \cite{cosme1}.
In principle, interaction strength, detunings, and hopping of few atoms in a double well can be controlled almost arbitrarily \cite{RevModPhys.80.885}. Putting many of these double wells in a row and measuring their independent contribution to the signal (see \cite{oberthaler14}) gives sufficient detection efficiency.

Finally, another important experimental panorama is that of magnetic systems: key examples are molecular nanomagnets \cite{nanomagnets} and nitrogen vacancies (N-V) implanted in solid crystals \cite{Doherty2013}.
The former are highly engineerable and effective isolation of few energy levels having pure-spin degrees of freedom can be achieved. Dynamical control can then be naturally implemented by means of time-dependent magnetic fields. N-V centers have the advantage of permitting typically longer coherence times. 

In general, the Hamiltonian corresponding to our model, Eq. \eqref{eq:hamiltonian}, may describe an effective spin-1 system. This can be represented, for instance, by a part of the energy spectrum of a molecular nanomagnet \cite{collison1}, or by the $S=1$ electronic ground state of negatively charged N-V centers in diamond \cite{kubo2010,Doherty2013}. The Hamiltonian is $ \hat H(t) = \hbar D \hat S_z^2 + g \mu_B \bm{B} \cdot \bm{\hat S}$, with $\hbar D>0$ representing the axial zero-field splitting parameter, $B_z(t)=\dot B_z t$ sweeping linearly in $t$ (to implement the LZ dynamics), $B_x$ constant in time and null $B_y$ component.
Direct mapping to the dimensionless Hamiltonian \eqref{eq:hamiltonian} can be given by the identification $\epsilon = \hbar D/g \mu_B B_x, \alpha = \hbar \dot B_z/g \mu_B B_x^2 , \tau = g \mu_B B_x t /\hbar$, $\Delta=1$, if the characteristic energy which defines the energy unit is chosen to be $g \mu_B B_x$. 
In this case, the implementation of the full control Hamiltonian of Eq. \eqref{eq:H1} is not obvious, due to the necessity of dynamically controlling the direct $[(13)]$ coupling element ($\Delta m=\pm 2$ states in terms of $\hat S_z$ eigenstates). Nonetheless, the approximate control based on two-level independent corrections (see Sec. \ref{separability}) could be implemented by modulating in time an additional $B_y(t)$ component, $\hat H^{(1)}_{\text{sep}}(t) = g \mu_B B_y \hat S_y$, with profile $B_y =  B_x \partial_t \theta_{\text{sep}}(t)/\sqrt{2}$ and $\partial_t \theta_{\text{sep}}(t)$  given in Eq. \eqref{sepprofileBy}.

We conclude this section with a discussion concerning typical experimental constraints. 
The TQD algorithm is by construction rather robust against experimental imperfections. For a detailed discussion of this point, in the context of the experimental realization of a single LZ superadiabatic control, see Ref. \cite{bason1}. 
Let us now concentrate on the specific three-level system under study, considering as a reference example a realization based on magnetic systems. We refer to the identification of the parameters of a magnetic system described in the previous paragraph.

There are two main aspects which should be discussed, namely, the realizability of the scheme of levels proposed (i.e., of the sequence of LZ avoided crossings; see Fig. \ref{fig:spec}) and the implementation of the control fields.
Concerning the scheme of levels, a typical experimental difficulty is the initial state preparation. This is not a problem in the case of magnetic realizations, especially due to the fact that we are interested in driving the ground state, as can be achieved by proper cooling of the system following well-established standard procedures.
Furthermore, a remarkable point is the fact that coherence timescales, which are usually one of the main experimental limitations, do not play a central role in our framework. 
In fact, if the control is efficient, the system remains always in the ground state, and therefore is only weakly affected by decoherence effects \cite{Boixo2012}.

The implementation of the necessary intensities of the magnetic fields should also be inspected. In the optimal regime of parameters, once $\hbar D$ is fixed to typical values ($\sim 10^{-3} \,\text{meV}$) by engineering of the system, the constant $B_x$ realizing the coupling between diabatic levels assumes reasonable values, $\sim 10^{-3} \, \text{T}$. Moreover, the $B_z$ field, growing linearly in order to implement the LZ dynamics, should vary at a rate of $\sim 10-100 ~\text{T\,s}^{-1}$ reaching intensities of $\sim 10^{-1} \, \text{T}$ at the beginning or at the end of the sweep.
 
For what concerns the realizability of the control field, as mentioned in the previous paragraph and in Sec. \ref{separability} after Eq. \eqref{sepprofileBy}, a direct implementation of the full control is difficult in the case of a pure spin-1 system, due to the necessity of controlling in time the $\{ \hat H\}_{13}$ matrix element. Regarding the $B_y$ field implementing the separated-control instead, maximal intensities assume values of $\sim 10^{-3}\, \text{T}$ for the regime of the other parameters considered. The typical timescales of variation for the peaks are sufficiently long (10-100 $\mu \text{s}$) allowing for a precise modulation.


\section{Conclusions}\label{sec:conclusion}
We have studied the application of the transitionless driving protocol to a three-level quantum system whose energy spectrum is characterized by a sequence of LZ avoided crossings. 
Analytical expressions for the control Hamiltonian have been found for the time instant $\tau=0$, and for the case of simultaneous crossings, $\epsilon=0$. A numerical investigation of the general cases allows one to discuss the shape of the elements of the control Hamiltonian. We find that these are typically given by sharp pulses centered at the crossing times. Their tails decay as power laws at long-times; $\tau^{-2}$ for ``directly induced" crossings while $\tau^{-4}$ for the ``indirect crossing." Such behaviors are supported by perturbative calculations. This indicates that it is not possible to obtain exact control by applying two-level corrections to each crossing separately, a feature which ultimately suggests that LZ crossings are never completely independent events.

Nonetheless, we have addressed the issue of separability of the control Hamiltonian in terms of single crossing corrections from a practical point of view, by investigating how accurately the evolution generated by the separated control fields can reproduce a truly adiabatic evolution, as a function of the intercrossing separation $\epsilon$. Our results predict that, for large $\epsilon$, the nonadiabatic transition probability scales as $\epsilon^{-2}$.  

Finally, possible experimental realizations of the system studied here were discussed. Our predictions turn out to be robust with respect to small imperfections making the spectrum asymmetric. This opens the route for experimental verifications based, e.g., on ultracold atoms or molecular magnets, as well as for further studies on more than three-level systems. The TQD algorithm, while being typically not applicable analytically for larger systems, can be implemented numerically, in principle, in nondegenerate quantum systems with any number of levels. The separated-control strategy can be applied to quantum systems whose spectra are characterized by sequences of well-separated avoided crossings, if the time scales of separation of the anticrossings permit one to act on each single anti-crossing independently, in the sense discussed in Sec. \ref{separability}.

\acknowledgments{Helpful discussions with Giuseppe Allodi and Riccardo Mannella are gratefully acknowledged.
S.C. and P.S acknowledge financial support from the PRIN Project 2015 No. HYFSRT of the Italian Ministry of Education and Research (MIUR). 
}



%

\end{document}